 \documentclass[final,5p,times,twocolumn]{elsarticle}

\usepackage{amssymb}
\usepackage{lineno}
\usepackage{graphicx}
\usepackage{subfigure}
\usepackage[usenames,dvipsnames]{color}
\usepackage{bm}% bold math
\usepackage{epsfig}
\usepackage{amsmath,amssymb,amsfonts}
\usepackage{pict2e}
\usepackage{epstopdf}
\usepackage{amssymb}
\usepackage{amsthm}

\begin{document}
	
	\begin{frontmatter}
		
		
		
		\title{Analysis of cyclical behavior in time series of stock market returns}
		
		
		\author{Djordje Stratimirovi\'c}
		
		\address{Faculty of Dental Medicine, University of Belgrade, Dr Suboti\'ca 8, 11000 Belgrade, Serbia}
		\address{Institute for Research and Advancement in Complex Systems, Zmaja od no\'caja 8, 11000 Belgrade, Serbia}
		
		\author{Darko Sarvan}
		\address{Faculty of Veterinary Medicine, University of Belgrade, Bulevar oslobodjenja 18, 11001 Belgrade, Serbia}
		
		\author{Vladimir Miljkovi\'c}
		\address{Faculty of Physics, University of Belgrade, P.O. Box 550, 11001 Belgrade, Serbia}
		
		\author{Suzana Blesi\'c}
		\address{Department of Environmental Sciences, Informatics and Statistics, Ca'Foscari University of Venice, Campus Scientifico, Via Torino 155, 30172 Mestre, Italy}
		\address{Institute for Research and Advancement in Complex Systems, Zmaja od no\'caja 8, 11000 Belgrade, Serbia}
		
		\begin{abstract}
			In this paper we have analyzed scaling properties and cyclical behavior of the three types of stock market indexes (SMI) time series: data belonging to stock markets of developed economies, emerging economies, and of the underdeveloped or transitional economies. We have used two techniques of data analysis to obtain and verify our findings: the wavelet transform (WT) spectral analysis to identify cycles in the SMI returns data, and the time-dependent detrended moving average (tdDMA) analysis to investigate local behavior around market cycles and trends.
			We found cyclical behavior in all SMI data sets that we have analyzed. Moreover, the positions and the boundaries of cyclical intervals that we found seam to be common for all markets in our dataset. We list and illustrate the presence of nine such periods in our SMI data. We report on the possibilities to differentiate between the level of growth of the analyzed markets by way of statistical analysis of the properties of wavelet spectra that characterize particular peak behaviors. Our results show that measures like the relative WT energy content and the relative WT amplitude of the peaks in the small scales region could be used to partially differentiate between market economies. Finally, we propose a way to quantify the level of development of a stock market based on estimation of local complexity of market's SMI series. From the local scaling exponents calculated for our nine peak regions we have defined what we named the Development Index, which proved, at least in the case of our dataset, to be suitable to rank the SMI series that we have analyzed in three distinct groups.
		\end{abstract}
		
		
	\end{frontmatter}
	
	%\linenumbers
	
	\section{Introduction}
	%\label{}
	This paper seeks to investigate the appearance of periodic and non-periodic cycles in the time series of stock market returns, and the contribution of cyclic behavior to the market efficiency and the distribution of stock indexes returns. Cycles in the economic data have been studied extensively \cite{ref1}, resulting in a number of stylized facts that characterize some cyclical or seasonal effects to financial time series \cite{ref2}. The study of cycles in economic data dates back to the early 1930s \cite{ref3}. Various techniques to measure seasonality have been widely applied, combining ideas from mathematics, physics, economics and social sciences. These efforts have resulted in research findings of, among other, intraday trading effects \cite{ref4}, weekend and/or three-day effects \cite{ref5}, intramonth effects \cite{ref6}, quarterly and annual cycles \cite{ref7}, and various multi-year cyclical variations in stock market index returns \cite{ref3, ref8}. A consensus of opinion on the nature, character, or the importance (to the overall market data behavior) of cyclic effects, however, has not been reached.
	
	Financial markets belong to a class of human-made systems exhibiting complex organization and dynamics, and similarity in behavior \cite{ref9}. Complex systems have a large number of mutually interacting parts that operate simultaneously at different scales, are often open to their environment, and self-organize their internal structure and dynamics, thus producing various forms of large-scale collective behaviors. The outputs of such systems, time series of records of their activity, display co-existence of collectivity and noise \cite{ref9.1}; the complexity of systems is reflected in datasets that exhibit a wealth of dynamic features, including trends and cycles on various scales \cite{ref1, ref3}. The tools to study such systems therefore cannot be analytical, but rather must be adapted to enable accurate quantification of their long-range order. In this sense, we have chosen to contribute to the debate about the existence, types, and importance of cycles in stock market data in two ways: by way of applying wavelet spectral analysis \cite{ref10} to study market returns data, and through the use of Hurst exponent estimation methods \cite{ref11} to study local behavior around market cycles and trends. The utility of our methods to estimate the scaling of financial time series has recently been confirmed \cite{ref11a} in an extensive overview of scientific time series data and analysis methods.  
	
	Firstly, we utilized wavelets to study cyclical consistency in time series of stock market indexes (SMIs). Wavelet analysis is appropriate for such a task; it was originally introduced to study complex signals \cite {ref12}. We use wavelet-based spectral analysis, which estimates the spectral characteristics of a time-series as a function of time \cite{ref13}, revealing how the different periodic components of a particular time-series evolve over time. It enables us to compare stock market index time series wavelet spectra from different economies, and to examine the similarities in contributions of cycles at various characteristic frequencies to the total energy spectrum. With this tool we can attempt to address the question of whether the complexity of a financial market is specifically limited to the statistical behavior of each SMI time series or parts of an SMI's series complexity can be attributed to the overall world market \cite{ref14}.
	
	We use the Hurst exponent estimation formalism, in a form of time-dependent detrended moving average analysis, to test the local character of cycles at various characteristic frequencies of SMI time series from different economies. In recent years, the application of the Hurst-exponent-based analyses has led many researchers to conclude that financial time series possess multi-scaling properties \cite{ref15, ref16, ref17}. In addition, these methods have allowed for the examination of local scaling around a given instance of time, so that the complex dynamical properties of various time series can be analyzed locally rather than globally \cite{ref17'}. In this paper, we aim to compare the local scaling of each cycle across stock markets and to find ways to classify various markets according to their cyclical behavior.
	
	We choose to analyze three types of SMI time series: data belonging to stock markets of developed economies, emerging economies, and of the underdeveloped or transitional economies. Previous and recent work by our group and others has demonstrated that SMI series exhibit scaling properties connected to the level of growth and/or maturity of the economy the stock market is embedded in \cite{ref15, ref18}. It has also been demonstrated that in emerging or transitional markets stock indexes do not fully represent the underlying economies \cite{ref15}, therefore we wanted to tailor our SMI study with this in mind and differentiate between underdeveloped (transitional) economies, emerging economies, and developed economies.
	
	Our study is structured as follows. In Sec. 2. we give a brief overview of the methodological background: the general framework of the wavelet transform (WT) spectral analysis and an introduction to the detrended moving average (DMA) method and its time-dependent variation (tdDMA). In Sec. 3. we present our dataset and the results of the usage of the WT framework to study the appearance and consistency of cycles across stock markets. In addition, in this section we present the results of investigation of statistical effects of the observed cyclical behavior on the WT spectral behavior of our SMI data. In Sec. 4. we list the results of the use of tdDMA on our SMI data and develop a quantitative indicator (that we have dubbed the 'Development Index'), which may help classify the level of development of a particular market according to the markets' local cyclical behavior. We end our paper with a list of conclusions and a few suggestions for future work in Sec. 5.
	
	\section{Methodological background}
		\label{wt}
		
		In this paper we use the wavelet transform power spectrum and the time-dependent detrending moving average approaches for data analysis. 
		
		The wavelet transform (WT) was introduced \cite{ref19, ref20, ref20a} in order to circumvent the Heisenberg uncertainty principle problem in classical signal analysis and achieve good signal localization in both time and frequency that a classical Fourier transform approach lacks. Namely, in WT the window of examination length is adjusted to the frequency analyzed: slow events are examined with a long window, whilst a shorter window is used for fast events. In this way an adequate time resolution for high frequencies and a good frequency resolution for low frequencies is achieved in a single transform \cite{ref10}.
		
		The continuous wavelet transform \cite{ref19,ref20} of a discrete sequence $R(k)$ is defined as the convolution of $R(k)$ with wavelet functions $\psi_{a,b}(k)$ in the following way:
		$$W(a,b)=\sum_{k=0}^{N-1}R(k)\psi^*_{a,b}{(k)}\>, \eqno (1)$$
		\noindent with $a$ and $b$ being the scale and translation-in-time (coordinate) parameters, $N$ the total length of the time series studied, and the asterisk stands for complex conjugate. In order to examine the existence of cycles in SMI data, we used the wavelet scalegrams (mean wavelet power spectra) $E_W(a)$, that are defined by 
		$$E_W(a)=\int
		W^{2}(a,b)db\>. \eqno (2)$$\\
		The scalegram $E_W(a)$ can be related \cite{ref21} to the corresponding Fourier power spectrum $E_F(\omega)$ via the formula
		$$E_W(a)=\int
		E_F(\omega)|\hat{\psi}(a\omega)|^2d\omega\>, \eqno
		(3)$$\\
		where the hat designates the Fourier transform, while $E_F(\omega)=|\hat{R}(\omega)|^2$. This formula implies that if the two spectra, $E_W(a)$ and $E_F(\omega)$, exhibit power-law behavior, then they should have the same power-law exponent $\beta$. The meaning of the wavelet scalegram is the same as that of the classical Fourier spectrum - it gives a contribution to the signal energy at a specific scale parameter $a$. We are thus able to view and estimate the peaks of wavelet spectra in the same way as one would approach this problem in Fourier analysis. In this paper, we find it convenient to use the standard set of Morlet wavelet functions as a wavelet basis for our analysis. The Morlet wavelet \cite{ref21, ref22} has proven to possess the optimal joint time-frequency localization \cite{ref14, ref23}, and can thus be used for detecting locations and spatial distribution of singularities in time series \cite{ref24}.
		
		In another approach, we employed the detrended moving average (DMA) technique \cite{ref24a} to study the general statistics of our SMI data. We use the variation of a standard DMA method that is introduced in \cite{ref26}. This technique calculates the centered detrended moving average (cDMA) function \cite{ref27} of the type
		$$\sigma_{cDMA}(n)=\sqrt{{1\over{N_{max}-n}}\sum_{i=\frac{n}{2}}^{N_{max}-\frac{n}{2}}(y_{n}(i))^2}\>,
		\eqno (4)$$\\
		where $y_{n}(i)$ are fluctuations around the moving average of a time series, calculated on a segment size $n\leq N$.
		
		By increasing the segment length $n$ the function $\sigma_{cDMA}(n) \equiv \sigma(n)$ increases as well. When the analyzed time series follows a scaling law (i.e. exhibits self-similarity over a range of time scales), the cDMA function is of a power-law type, that is, $\sigma(n)\propto\ n^{H}$, with $0 \leq H \leq 1$. Scaling exponent $H$ is usually called the Hurst exponent of the series \cite{ref25}. In the case of short-range data correlations (or no correlations at all) $\sigma(n)$ behaves as $n^{1/2}$. For data with power-law long-range autocorrelations one may expect that $H > 0.5$, while in the long-range negative autocorrelation case we have $H < 0.5$. When scaling exists, the exponent $H$ can be related to the WT power spectrum exponent $\beta$ through the scaling relation \cite{ref28} $H = (\beta + 1)/2$.
		
		In order to inspect local cyclical behavior of our SMI series, we applied the time-dependent DMA algorithm (tdDMA) \cite{ref17'} to the subset of data in the intersection of the SMI signal and a sliding window of size $N_{s}$, which moves along the series with step $\delta_{s}$. The scaling exponent $H$ is calculated for each subset and a sequence of local, time-dependent Hurst exponent values is obtained. The minimum size of each subset $N_{min}$ is defined by the condition that the scaling law $\sigma(n)\propto\ n^{H}$ holds in the subset, while the accuracy of the technique is achieved with appropriate choice of $N_{min}$ and $\delta_{min}$ \cite{ref30}. We have chosen windows of up to $N_{s} = 1000$, with the step $\delta_{s} = 1$ for our tdDMA algorithm. 
		
		\section{Data and results}
		\subsection{Stock market data studied}
		\label{data}
		
		In this paper, we investigate data from the following stock markets: the New York Stock Exchange NYSE index, the Standard \& Poor's 500 (S\&P500) index, the UK FTSE 100 index, the Tokyo Stock Exchange NIKKEI 225 index, the French CAC 40 index, and the German Stock Market DAX index, which we consider developed economies; the Shanghai Stock Exchange SSE Composite index, the Brazil Stock Market BOVESPA index, The Johannesburg Stock Exchange JSE index, the Turkey Stock Market XU 100 index, the Budapest Stock Exchange BUX index, and the Croatian CROBEX index, which we consider emerging economies; the Tehran TEPIX index, the Egyptian Stock Market EGX 30 index, and the indexes of the developing economies in the Western Balkans - the Belgrade Stock Exchange BELEXline index, the Montenegrin MONTEX 20 index, the SASX 100 index of the market of Bosnia and Herzegovina and the BIRS index of Bosnian entity Republic of Srpska, representing markets of underdeveloped economies. Table 1 lists general characteristics of the SMI time series which we have analyzed; depending mainly on the market development level, they are of varying duration.
		
		\begin{table}[htbp]
			\caption{General characteristics of the SMI time series analyzed in this paper.}
			\label{tab:1}
			\resizebox{0.7\textwidth}{!}{\begin{minipage}{\textwidth}
					\begin{tabular}{llllll}
						\hline\noalign{\smallskip}
						SMI name (economy) & Recording period & Total days $N$\\
						\noalign{\smallskip}\hline\noalign{\smallskip}
						BELEXline (Serbia) & October 1, 2004 - December 31, 2014 & 2584\\
						SASX 10 (Bosnia and Herzegovina) & June 2, 2005 - February 11, 2015 & 2255\\
						BIRS (Republic of Srpska) & May 15, 2005 - February 10, 2015 & 2303\\
						TEPIX (Iran) & February 14, 2010 - February 10, 2015 & 1205\\
						MONTEX 20 (Montenegro) & May 1, 2004 - February 10, 2015 & 2745\\
						EGX 30 (Egypt) & January 1, 1998 - February 11, 2015 & 4179\\
						BOVESPA (Brasil) & April 27, 1993 - January 14, 2015 & 5383\\
						JSE (South Africa) & June 5, 2006 - February 11, 2015 & 2174\\
						SSE (China) & December 19, 1990 - December 5, 2014 & 6142\\
						CROBEX (Croatia) & September 2, 1997 - February 10, 2015 & 4323\\
						XU 100 (Turkey) & June 2, 2003 - February 10, 2015 & 2922\\
						BUX (Hungary) & April 1, 1997 - February 10, 2015 & 4465\\
						FTSE 100 (UK) & March 1, 1984 - February 10, 2015 & 8109\\
						CAC 40 (France) & March 1, 1990 - February 10, 2015 & 6320\\
						NIKKEI 225 (Japan) & April 1, 1984 - December 18, 2014 & 7625\\
						NYSE (USA) & March 1, 1966 - February 10, 2015 & 12365\\
						DAX (Germany) & November 26, 1990 - February 10, 2015 & 6131\\
						S\&P 500 (USA) & March 1, 1950 - February 10, 2015 & 16383\\
						\noalign{\smallskip}\hline
					\end{tabular}
				\end{minipage}}
			\end{table}
			
			The variables studied in our paper are the daily price logarithmic returns that are defined as
			$$R(t) = logS(t+\Delta t)-logS(t)=log({{S(t+\Delta t)}\over {S(t)}})\>,
			\eqno (5)$$\\
			\noindent where $S(t)$ is the closure price of the stock market index at day $t$, and the lag period $\Delta t$ is a time interval of recording of index values $S(t)$. All of the analyzed time series of prices on the stock markets $S(t)$ are publicly available (from the official web-sites of the markets in question, or from the Yahoo Finance Database), and are given in local currencies. The values of the SMI data are listed only for trading days -- that is, they are recorded according to the market calendar, with all weekends and holidays removed from datasets.
			
		\subsection{Wavelet spectra of stock market data}
		
\begin{figure}[h]
		\includegraphics[scale=0.5]{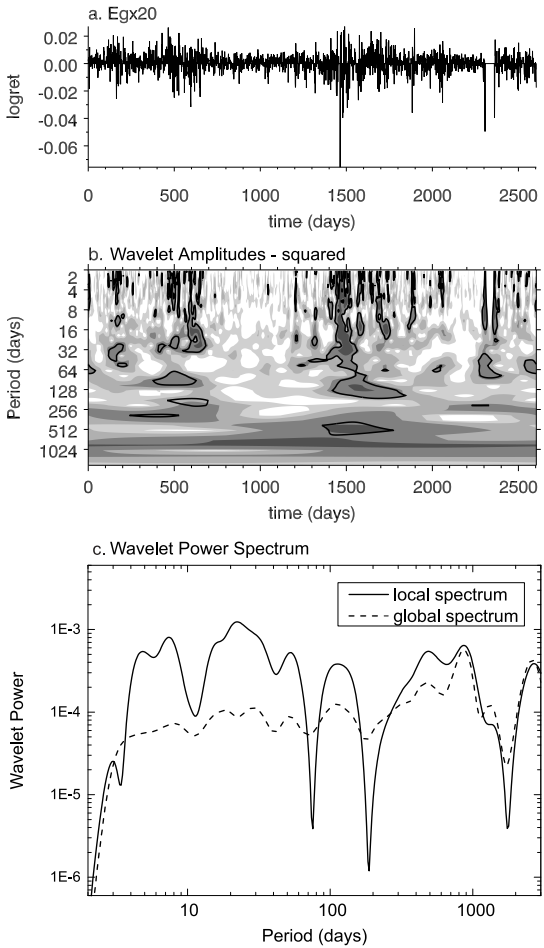}
		\caption {An example of a significance test for peaks in EGX 30 wavelet power spectrum. (a) Raw data; (b) The local wavelet power spectrum. 
			The contour levels are chosen so that $ 75\% $, $ 50\% $, $ 25\% $, and $ 5\% $ of the wavelet power is above each level, respectively. Black contour is the $ 10\% $ significance level, using the global wavelet as the background spectrum; (c) Comparison of the local wavelet power spectrum, calculated at 1500 points, with the global wavelet spectrum for the same sat of data. Significant peaks appear above the global spectrum.} 
		\label{Fig1}		
	\end{figure}

		We have calculated WT power spectra for all our SMI series, and for all the periods (durations) these data series were available to us. We took into consideration only the values of the WT spectra between the minimum time scale of $a=1$ and the statistically meaningful maximum time scale \cite{ref25} of $a=N/5$, and searched for characteristic peaks (local maxima) within those limits. In order to be sure that the peaks that we have obtained in such a way are not artefacts of WT method used, we have additionally performed a test of statistical significance for each peak, using the tool kit described in \cite{ref21.1} and ready-to-use software available online at \cite{ref21.2}. In order to assess the significance of each peak, we compared them against the background global wavelet spectrum that they belong to. We have first calculated the local WT spectra of each SMI series and have searched for WT coefficients with a $10\%$ significance value. We have then calculated the local WT spectra on the time scales that show existence of broad WT significance over many periods. The peaks that appeared above global spectrum were then used as significant for further analysis. Figure \ref{Fig1}. depicts the way this significance test was done, on an example of the EGX 30 time series. The choice to use global wavelet spectra as the background against which the significance of peaks was tested was guided by the fact that the SMI time series are products of a complex system that result from interactions of many constituents acting on different time scales. The SMI time series are thus mixtures of noise components from different inputs involved in the process \cite{ref21.3}; this fact renders it implausible to compare peaks from SMI wavelet spectra against any particular noise background other than the signal itself \cite{ref21.4}.   
		
		We found multiple peaks in all SMI series of our dataset. Moreover, the peaks we found show commonality across the dataset, that is, if they exist peaks appear at relatively similar positions (characteristic times). The following common characteristic peaks, or rather characteristic cycle periods around characteristic peaks were identified by our analysis: a working-week cycle  (or a 5-day peak), a one-week cycle (or a 7-day peak), a two-week cycle (or a 14-day peak), a monthly cycle (or a 30-day peak), a quarterly cycle (or a 90-day peak), a 4- to 5-month cycle (or a 150-day peak), a semi-annual cycle (or a 6- to 7-month peak), an annual cycle (or a 360-day peak), and a bi-annual (or a 600-days) multi-year cycle. The peaks that we found in each individual SMI series are listed in Table 2. The dissimilarities between SMI records from different economies that we observed occur only in the lack of a spectral peak (see Table 2), or a slight lack of synchronization of a particular peak position (that is, we found that peaks are not positioned at exactly the same time instances in all the SMI series analyzed, which prompted us to introduce the notion of a peak or a cycle interval). In Table 2 the cycles and the cycle intervals are given in real days (recalculated from trading days that comprise our raw data).  
		
		\begin{table}[htbp]
			\caption{An overview of cycles in SMI time series identified by the wavelet spectrum analysis.}
			\label{tab:2}
			\smallskip
			\resizebox{0.6\textwidth}{!}{\begin{minipage}{\textwidth}
					
					\begin{tabular}{|l|c|c|c|c|c|c|c|c|c|}
						%\multicolumn{10}{c}{\bf{relative energy content under the peaks}}\\
						\hline
						peak interval number & I & II & III & IV & V & VI & VII & VIII & IX \\
						\hline
						peak at (days) & 5 & 7 & 14 & 30 & 90 & 150 & 210 & 360 & 600 \\
						\hline
						interval length (days) & 2-6 & 6-10 & 10-25 & 25-60 & 60-110 & 110-190 & 190-250 & 250-450 & 450-900 \\
						\hline
						BELEXline & x & x &  & x &  & x & x &  & x \\
						SASX 10   & x & x & x & x & x & x &  & x & x \\
						BIRS      & x & x & x & x &  & x &  & x &  \\
						TEPIX     &  & x & x & x & x & x &  & x & x \\
						MONEX 20  & x & x & x & x &  & x & x & x & x \\
						EGX 30    & x & x & x & x & x & x & x &  & x \\
						BOVESPA   & x & x & x & x & x & x &  & x & x \\
						JSE       & x & x & x & x &  & x & x & x &  \\
						SSE       & x &  & x & x & x & x & x &  & x \\
						CROBEX    & x & x & x & x & x & x & x & x & x \\
						XU 100    & x &  & x & x & x & x &  & x & x \\
						BUX       & x & x & x & x & x & x &  & x & x \\
						FTSE 100  & x & x & x & x & x & x & x &  & x \\
						CAC 40    & x & x & x & x & x & x &  & x &  \\
						NIKKEI 225    & x & x & x & x & x &  &  & x & x \\
						NYSE      & x &  & x & x & x & x &  & x & x \\
						DAX       & x & x & x & x & x & x & x & x & x \\
						S\&P 500  & x & x & x & x & x & x &  & x & x \\
						\hline
					\end{tabular}
				\end{minipage}}
			\end{table}
			
			The examples of detected peaks and subsequently defined peak intervals are given in Figs. 2 and 3.
			
			\begin{figure}[h]
				\includegraphics[scale=0.35]{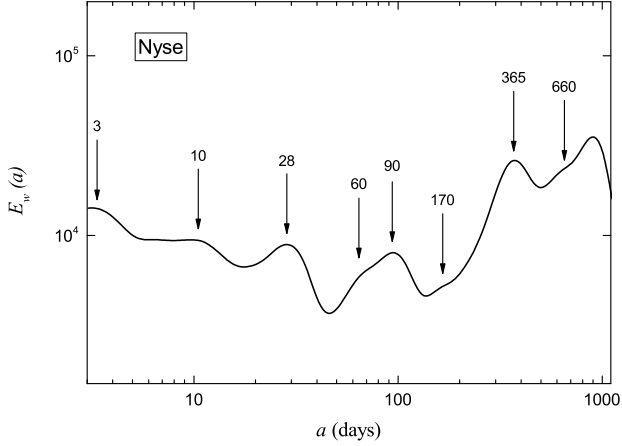}%
				\caption{An example of detected peaks in the time series of NYSE SMI data. \label{Fig2}}%
			\end{figure}
			
			\begin{figure}[h]
				\includegraphics[scale=0.35]{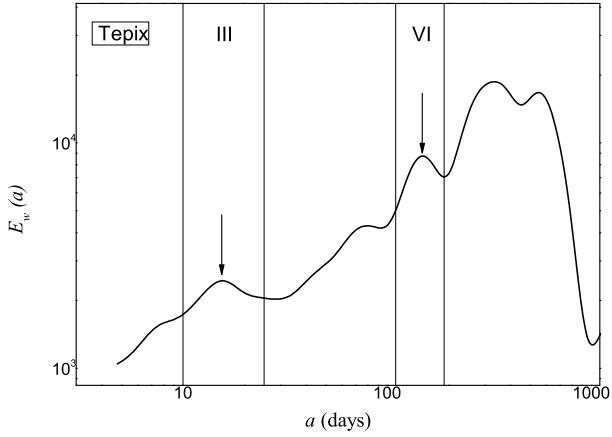}%
				\caption{An illustration of positioning of two peak intervals in a time series of TEPIX SMI data. \label{Fig3}}%
			\end{figure}
			
			\subsection{Statistical characterization of WT spectra of stock market data}
			
			In order to be able to compare and characterize the obtained wavelet spectra of our stock market data, we have calculated relative energy content and relative amplitude of all the regions (listed in Table 2) under characteristic peaks in all our data series. The relative energy content of the $i$-th peak in a WT power spectrum is defined \cite{ref10} as:
			$$e_{Wi}(s_{i1},s_{i2})=\frac{E_{i}(s_{i1},s_{i2})}{E_{total}} \>, \eqno
			(6)$$
			where $E_{i}(s_{i1},s_{i2})$ represents the average energy content of the period surrounding the $i$-th peak:
			$$E_{i}(s_{i1},s_{i2})=\frac{1}{t}\int_{0}^{t}\int_{1/2\pi s_{i2}}^{1/2\pi s_{i1}}\frac{1}{a^{2}}|W(a,b)|^2 da db \>, \eqno
			(7)$$
			and $E_{total}$ is the total energy content of the WT spectrum of the stock market series analyzed. The energy content is a physical quantity behind a WT power spectrum, so it represents it's natural characteristic. Similarly, the relative amplitude of the spectral band under the $i$-th peak is defined as:
			$$a_{Wi}(s_{i1},s_{i2})=\frac{A_{i}(s_{i1},s_{i2})}{A_{total}} \>, \eqno
			(8)$$
			with
			$$A_{i}(s_{i1},s_{i2})=\frac{1}{t}\int_{0}^{t}\frac{1}{s_{i2}-s_{i1}}\int_{1/2\pi s_{i2}}^{1/2\pi s_{i1}}\frac{1}{a^{2}}W(a,b) da db \>, \eqno
			(9)$$
			its average amplitude, and $A_{total}$ the total amplitude of the WT power spectrum of the stock market series of interest. The amplitude of the WT power spectrum depends \cite{ref10} on the variability of the frequency (scale) band analyzed - the more constant the frequency, the higher the amplitude.
			
			We calculated the relative energy contents $e_{Wi}$ and the relative amplitudes $a_{Wi}$ for all the obtained peaks in all the analyzed WT spectra. We then performed statistical analysis of three groups of data - those belonging to the developed economies, the emerging markets, and the underdeveloped economies. We first performed the Shapiro-Wilk test for normality of distributions within these three data groups. If normality of distributions existed within our datasets, we performed the one-way ANOVA test to compare our sample means, with the significance level of $p<0.05$. If the ANOVA test confirmed the existence of differences of means, the average means for all three groups of data was compared using the Bonferroni method. If, however, the Shapiro-Wilk test did not confirm the existence of normality of distributions within our dataset, we performed the Kruskal-Wallis ANOVA test to compare the means, with the significance level of $p<0.05$. If the Kruskal-Wallis ANOVA test confirmed the existence of differences in the groups' means, the comparison of average means for all three groups of data was done using the Wilcoxon Mann-Witney method.
			
			Table 3. lists the calculated average values of relative energy content $e_{Wi}$ and the relative amplitudes $a_{Wi}$ of all the peaks for the three SMI groups. The statistically significantly different values between the groups for each of the peaks are marked in bold - if only one value is bolded, then it differs from the other two market groups in a peak group; if two values are bolded they differ mutually; and if all three values have been bolded then all the three market groups' values differ from each other.
			
			\begin{table}[htbp]
				\caption{Values of relative energy contents and relative amplitudes under WT peaks. The statistically significantly different values between the groups for each of the peaks are marked in bold. When one value is bolded, then it differs from the other two market groups in a peak group. If two values are bolded they differ mutually, and if all three values are bolded then all the three market groups' values differ from each other.}
				\label{tab:3}
				\smallskip
				\resizebox{0.6\textwidth}{!}{\begin{minipage}{\textwidth}
						
						\begin{tabular}{|l|c|c|c|c|c|c|c|c|c|}
							\multicolumn{10}{c}{\bf{relative energy content under the peaks}}\\
							\hline
							peak at (days) & 5 & 7 & 14 & 30 & 90 & 150 & 210 & 360 & 600 \\
							\hline
							underdeveloped & \bf{0.0004} & \bf{0.0006} & \bf{0.0028} &\bf{ 0.0055 }& \bf{0.012} & 0.0087 & 0.024 & 0.051 & \bf{0.34} \\
							
							emerging & \bf{0.0017} & 0.0022 & \bf{0.0079} & 0.015 & 0.017 & 0.016 & 0.039 & 0.09 & 0.45 \\
							
							developed & \bf{0.0032} & 0.0033 & \bf{0.012} & 0.019 & \bf{0.023} & 0.014 & 0.038 & 0.057 & 0.39 \\
							\hline\noalign{\smallskip}
							\multicolumn{10}{c}{\bf{relative amplitudes under the peaks}}\\
							\hline
							peak at (days) & 5 & 7 & 14 & 30 & 90 & 150 & 210 & 360 & 600 \\
							\hline
							underdeveloped & \bf{0.0009} & \bf{0.0012} & \bf{0.0061} & \bf{0.0098} & \bf{0.019} & \bf{0.016} & 0.037 & 0.063 & \bf{0.32} \\
							
							emerging & 0.002 & 0.0025 & 0.011 & 0.017 & 0.023 & \bf{0.024} & 0.046 & 0.081 & \bf{0.37}\\
							
							developed & 0.0026 & 0.003 & 0.013 & 0.019 & \bf{0.026} & 0.022 & 0.046 & 0.066 & 0.34 \\
							\hline
						\end{tabular}
					\end{minipage}}
				\end{table}
				
\begin{figure}[h]
					\includegraphics[scale=0.34]{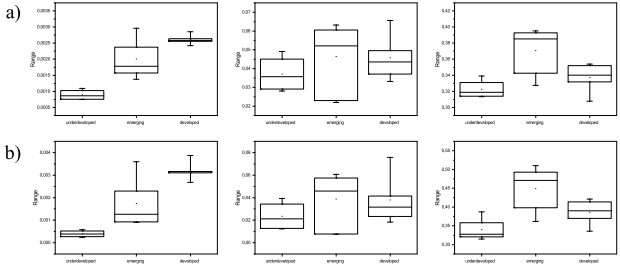}%
					\caption{Results of the statistical analysis of differences between average values of: a) the relative energy content $e_{Wi}$ and b) the relative amplitudes $a_{Wi}$ for all three market groups. Results are depicted for the three peak regions - a small scale region surrounding the peak at 5 days, a mid-scale region surrounding the peak at 150 days, and a large scale region surrounding the peak at 600 days. Squares enclose the $75\%$ of the values within the SMI group, while the error bars depict the maximum and the minimum value within the same group.\label{Fig4}}%
				\end{figure}

				Our results are also illustrated in Figure \ref{Fig4}, where average values of the relative energy content $e_{Wi}$ and the relative amplitudes $a_{Wi}$ for all three market groups, and in three peak regions - a small scale region surrounding the peak at 5 days, a mid-scale region surrounding the peak at 150 days, and a large scale region surrounding the peak at 600 days, are depicted. Table 3 and Figure \ref{Fig4} show that in the small scales regions (peaks of up to 90 days) the values of both the relative energy contents $e_{Wi}$ and the relative amplitudes $a_{Wi}$ under the spectral peaks for the underdeveloped markets are smaller than the values for the two other groups in a clear, statistically significant manner. Even more so, the values of the relative energy content $e_{Wi}$ for the small scale peaks at 5 days and at 14 days are statistically different for all three market groups. For the peaks at lager scales (peaks at 150 days and more), the behavior of underdeveloped markets data does not differ from the other two groups, except in the case of a large scale region of the peak at 600 days. It seems, therefore, that the transitional markets do not follow the same behavioral pattern as the markets of emerging or developed economies at short time scales of days, weeks, and several months. Our results thus show that measures like $e_{Wi}$ and $a_{Wi}$ for the peaks in the small scale regions could be used for partial differentiation between market economies.

				\section{Time dependent analysis of stock market data}
				\label{tddma}
				
\begin{figure}[h]
					\includegraphics[scale=0.35]{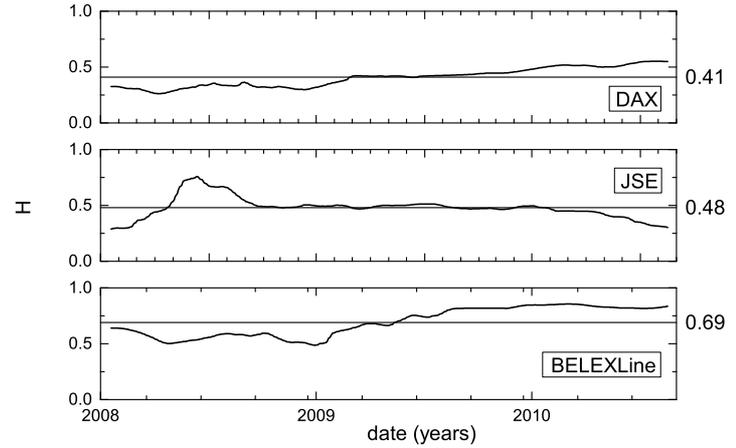}%
					\caption {
						An illustration for the calculated tdDMA values
						(local Hurst scaling exponents) in the case of the BELEXline SMI series (representing the markets of underdeveloped
						economies), the JSE SMI series (representing the emerging
						markets), and the DAX series (representing the markets of developed economies). The calculated tdDMA values are given
						for a time section from year 2008 to year 2011. Horizontal
						solid lines mark the values of the average (or global) Hurst scaling exponents for the same time period. Here, a moving window of $N_{s} = 1000$ and the step $\delta_{s} = 1$ were used. The error bars are not depicted here; for the estimation of errors to local Hurst exponents see \cite{ref25}.\label{Fig5}
					}%
				\end{figure}

				In order to gain another insight into the local complexity of our SMI data, and obtain a possibility to improve our ability to quantitatively distinguish the three groups of SMI data we use, we have applied the time-dependent detrended moving average (tdDMA) algorithm to all our SMI series. In Figure \ref{Fig5} we give an example of the calculated tdDMA values for the three randomly selected representatives of SMI market groups, in a time interval from year 2008 to year 2011, for a moving window of $N_{s} = 1000$, and the step $\delta_{s} = 1$.
				
				In an attempt to quantify the local behavior of SMI data and ultimately compare the efficiency of our stock markets, we have constructed the SMI Hurst vectors $h^{\alpha}$, where each coordinate $h_{i}^{\alpha}$ corresponds to the value of local Hurst exponent for a selected peak interval (that includes and borders each peak). Our calculations were performed on nine intervals that separate nine market peaks (listed in Table 2 and illustrated in Figs. 2 and 3), marked by index $i$ ($i=1 ... 9$), while $\alpha$ counts the SMI series. From all these values we have built the Hurst reference SMI vector $m$, where $m(i)$ represents the mean value of $h_{i}^{\alpha}$ for each coordinate (peak) $i$ across all the SMIs in the dataset. The Hurst reference SMI vector is thus defined as:
				$$
				m_i={1\over n} \sum_{\alpha=1}^{n}h_i^{\alpha}\>,
				\eqno (10) \label{Eq_mi}
				$$
				for $n=18$ different SMI indexes in our dataset. We have looked into how values of the reference vector $m_i$ are changing with the addition of new SMI data (markets), and in the case of our dataset this change becomes insignificantly small for $n>15$. The values of calculated Hurst vectors and Hurst reference vectors are listed in Table 4.
				
				\begin{table}[htbp]
					\caption{Hurst vectors ${h}^{\alpha}_i$ and the Hurst reference vector $m_{i}$ of stock market time series. Here, index $i$ numbers peak areas, while the index $\alpha$ marks stock markets.}\label{Tab_4}
					\resizebox{0.75\textwidth}{!}{\begin{minipage}{\textwidth}
							\smallskip
							\begin{tabular}{|cl|c ccc c c c c c|}
								
								%\multicolumn{10}{|l|c c c c c|c|c|c|c|}{peaks at (days) & 5 & 7 & 14 & 30 & 90 & 150 & 210 & 360 & 600 } \\
								
								\cline{3-11}
								\multicolumn{2}{c}{} & \multicolumn{9}{ |c| }{peak at (days)} \\
								\hline
								$\alpha$& &5&7&14&30&90&150&210&360&600 \\
								\hline
								1&$BELEXline$ & 0.36 & 0.62 & 0.59 & 0.67 & 0.71 & 1.01 & 0.90 & 0.68 & 0.59 \\
								2&$SASX 10$   & 0.38 & 0.48 & 0.47 & 0.60 & 0.63 & 0.89 & 1.01 & 0.90 & 0.80 \\
								3&$BIRS$      & 0.37 & 0.54 & 0.56 & 0.54 & 0.57 & 0.77 & 1.02 & 0.78 & 0.76 \\
								4&$TEPIX$     & 0.38 & 0.63 & 0.61 & 0.72 & 0.71 & 0.62 & 0.59 & 0.69 & 0.92 \\
								5&$MONEX 20$  & 0.37 & 0.53 & 0.50 & 0.56 & 0.51 & 0.54 & 0.70 & 0.81 & 0.93 \\
								6&$EGX 30$    & 0.38 & 0.58 & 0.52 & 0.49 & 0.73 & 0.85 & 0.77 & 0.72 & 0.43 \\
								7&$BOVESPA$   & 0.37 & 0.46 & 0.39 & 0.49 & 0.57 & 0.72 & 0.71 & 0.71 & 0.69 \\
								8&$JSE$       & 0.38 & 0.51 & 0.51 & 0.55 & 0.36 & 0.48 & 0.93 & 0.98 & 0.72 \\
								9&$SSE$       & 0.34 & 0.53 & 0.51 & 0.55 & 0.57 & 0.58 & 0.44 & 0.60 & 0.73 \\
								10&$CROBEX$    & 0.36 & 0.48 & 0.50 & 0.57 & 0.61 & 0.65 & 0.52 & 0.50 & 0.58 \\
								11&$XU 100$    & 0.37 & 0.52 & 0.47 & 0.57 & 0.49 & 0.56 & 0.56 & 0.70 & 0.55 \\
								12&$BUX$       & 0.37 & 0.46 & 0.44 & 0.47 & 0.45 & 0.50 & 0.56 & 0.64 & 0.47 \\
								13&$FTSE 100$  & 0.38 & 0.50 & 0.44 & 0.53 & 0.47 & 0.49 & 0.34 & 0.29 & 0.22 \\
								14&$CAC 40$    & 0.37 & 0.47 & 0.42 & 0.44 & 0.47 & 0.53 & 0.43 & 0.48 & 0.68 \\
								15&$NIKKEI 225$    & 0.36 & 0.47 & 0.43 & 0.49 & 0.53 & 0.58 & 0.46 & 0.50 & 0.56 \\
								16&$NYSE$      & 0.39 & 0.53 & 0.47 & 0.49 & 0.45 & 0.53 & 0.50 & 0.51 & 0.57 \\
								17&$DAX$       & 0.36 & 0.49 & 0.44 & 0.45 & 0.47 & 0.55 & 0.58 & 0.59 & 0.56 \\
								18&$S\&P 500$  & 0.38 & 0.50 & 0.47 & 0.49 & 0.47 & 0.53 & 0.52 & 0.55 & 0.52 \\
								\hline
								& $m_i$ & 0.37 & 0.51 & 0.49 & 0.54 & 0.54 & 0.62 & 0.64 & 0.63 & 0.61 \\
								\hline
							\end{tabular}
						\end{minipage}}
					\end{table}
					
\begin{figure}[h]
						\includegraphics[scale=1.2]{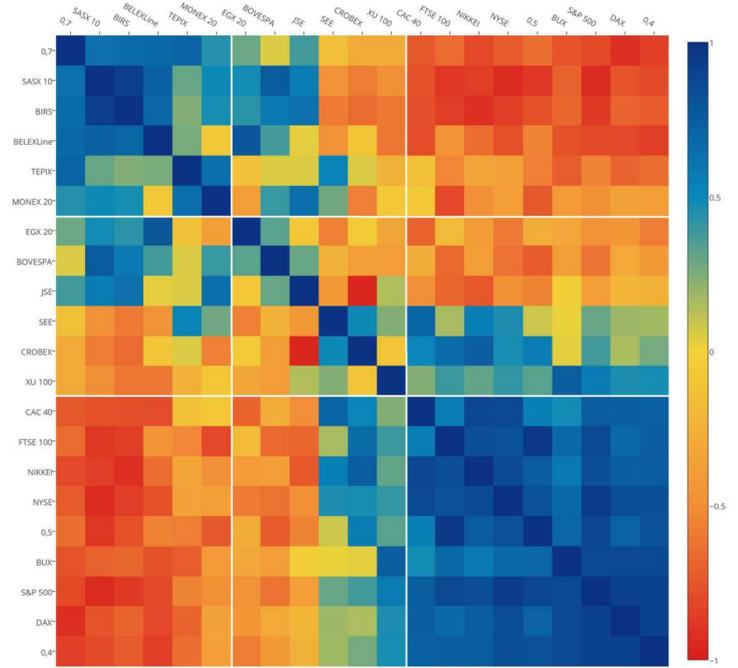}%
						\caption {
							Graphical representation of similarity, or similarity matrix, of relative Hurst unit vectors $s^\alpha_i$. Positive similarities of market's Hurst unit vectors are given in shades of blue (for ${\mathcal H}^{\alpha\beta}>0$), while negative similarities are depicted in shades of red (${\mathcal H}^{\alpha\beta}<0$). Horizontal and vertical white lines mark, from left to right, borders between groups of underdeveloped, emerging, and developed markets. \label{Fig6}
						}%
					\end{figure}

					From the two vectors $h^{\alpha}$ and $m$ we have calculated the relative SMI Hurst unit vectors $s^\alpha$ that we have defined as:
					$$s_i^\alpha=\frac{h_i^\alpha-m_i}{\sqrt{\sum_{i=1}^{n}(h_i^\alpha-m_i)^{2}}} \>.
					\eqno (11)$$
					Defined in such a way, the unit vectors $s^\alpha$ give us the information on the direction of difference between the Hurst vector ${h}^{\alpha}$ for each market and the Hurst reference vector $m$. We were hoping that this standard scoring will, to a certain accuracy, mark the overall financial status (i.e. development) of the markets in the dataset that we use. However, in the case of our dataset, the distance of representative points $s^\alpha$ from the Hurst reference point did not provide us with any relevant additional information about the market development or efficiency. This can be demonstrated through the use of the cosine similarity, a scalar Euclidean product of two $s^\alpha_i$ vectors that can quantify the level of similarity of positions of $s^\alpha$ for different SMI series. Scalar products of $s^\alpha_i$ are defined as:
					$${\mathcal H}^{\alpha\beta}=\sum_{i=1}^{p}s_i^\alpha s_i^\beta \>,
					\eqno (12)$$\\
					where $\alpha$ and $\beta$ count SMI series ($\alpha,\beta\in\{1,2, ..., 18\}$), while $p=9$ numbers peaks (peak regions). We have arranged and graphically presented values of these scalar products in Figure \ref{Fig6} for all our data and for three artificially produced time series with the values of H equal to 0.4, 0.5, and 0.7 in all of the analysed peak regions. These new series were added to serve as visual guides that separate different kinds of long-range behavior (that is, long-range anticorrelated behavior in the case H=0.4, uncorrelated behavior in the case H=0.5, and long-range correlated behavior for H=0.7). Figure \ref{Fig6} displays the existence of two separate block matrices that differentiate strong similarity within the group of underdeveloped markets (upper left corner) and within the group of developed markets (lower right corner), and strong dissimilarity inversely. Additionally, in Figure \ref{Fig6} the existence of a third market group is visible, that does not belong neither to developed nor to underdeveloped type. Members of this third group - the emerging markets - are weakly similar to both other two groups and within its own group, and show random unpredictable strong similarities with some members (markets) in the developed or the underdeveloped market group. This inability to 'look alike' differentiates emerging markets in Figure \ref{Fig6}, but not in a clear clustering way. 
					
					\subsection{The Development Index}
					
					In order to try to find a unique Hurst indicator that would be able to discern all our three categories of market development we have decided to define a (prefered) direction of development in markets indexes Hurst space, and then project unit vectors $s^\alpha$ onto that direction. We have decided to define this prefered direction as a direction of development, so that the projection of unit vectors from our developed markets group onto this direction will always be positive (this is why we have dubbed projections of unit vectors onto this pre-defined direction the Development Index). We have chosen the unit vector of development in the Hurst space in a following way: 
					$$
					e_{i}={\Delta h_i-m_i\over\sqrt{\sum_{i=1}^p(\Delta h_i-m_i)^2}}\>,
					\eqno (13) \label{ni}
					$$
					with $\Delta h_i=-I_i$, where $I_i$ stands for the $p$-vectors made of all unit components. 
					
\begin{figure}[h]
						\includegraphics[scale=0.55]{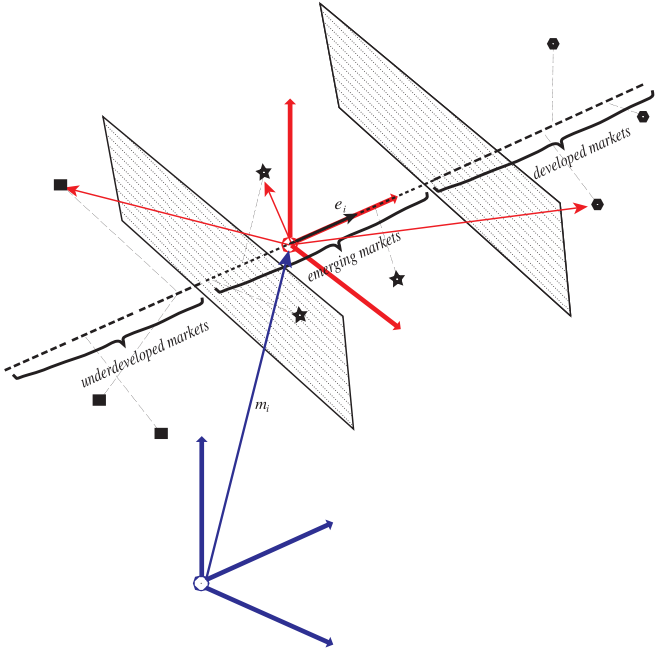}%
						\caption {Hurst parameter space represented by p-vectors in two reference spaces: in the general and the 'developed' (or relative) Hurst reference space (depicted by blue and red lines, respectively). The relative Hurst reference space is defined by the Hurst reference vector $m_i$, while the direction of the main axes is given by the unit vector of the direction of developemnt $e_i$. The  Development index $\Pi_{e_i}(s_i)$ is calculated as a projection of Hurst unit vectors ${s}^{\alpha}_i$ onto the $e_i$, which is directed to a portion of Hurst space where representative points of developed markets are grouped.   \label{Fig7}  }%
					\end{figure}

					In the case of our dataset, the values of this new vector's components have not significantly changed with the addition of new SMI data to dataset for $n>15$ ($n$ being the number of markets in the dataset analyzed). The relations in Eq. 13 led us to the value of $e_i$ for our dataset of $n=18$ stock market indexes:
					$$
					e_i= (- 0.19, - 0.40, - 0.37, - 0.45,- 0.45, - 0.57, - 0.60, - 0.59, - 0.56)\>,
					\eqno (14)
					$$
					with the error for each component $i$ being $\delta n_i=10^{-2}$. We have then calculated the Development Index (DI) as a projection of Hurst unit vectors onto this direction of development:
					$$
					\Pi_{e_i}(s_i)=\sum_i^ps_i e_i\>.
					\eqno (15)
					$$
					Graphical illustration of these projections is given in Figure \ref{Fig7}.
					
					Values of DI for markets in our dataset are given in Table \ref{Tab_5}. It is visible from Table \ref{Tab_5} that the three market categories (underdeveloped, emerging, and developed markets) can be differentiated by this order parameter. We have decided to define the borders that separate our three market categories using the following phenomenological arguments: since the values of the Hurst vectors $s^\alpha$ and their similarity that we have calculated point to the existence of two distinct groups that are well clustered (underdeveloped and developed markets), divided by a group of SMI time series that transitions between these two groups (the emerging markets), we used the symmetry principle to define a border between the group of developed and emerging markets at $\Pi_{c1}=\lvert \Pi\rvert_{max}/2 \pm 0.01$, and a border between the underdeveloped and emerging markets at $\Pi_{c2}=-\lvert \Pi\rvert_{max}/2 \pm 0.01$ (for our dataset, $\lvert \Pi\rvert_{max}=1.36$). Based on this criterion, in the case of our dataset, the Egyptian stock market index EGX30 would be classified as an emerging market, rather than an underdeveloped market as we initially assumed, while the Hungarian BUX index would classify as developed rather than the emerging market SMI.
					
					\begin{table}[htbp]
						\caption{Classifications of stock markets into clusters according to their maturity or development.\label{my-label}}
						\resizebox{0.8\textwidth}{!}{\begin{minipage}{\textwidth}
								\begin{tabular}{c c c c c c}
									\hline
									\multicolumn{1}{c}{} & \multicolumn{5}{ c }{underdeveloped markets} \\
									\hline
									\multicolumn{1}{c}{} & SASX10   & BIRS    & BELEXLine   & TEPIX   &  MONEX20 \\    
									
									$\quad\Pi_{e_i}(s_i)$ & -1.20      & -1.14      & -1.14   & -0.97    & -0.73 \\  
								\end{tabular}
								
								\begin{tabular}{c c c c c c c}
									\hline
									\multicolumn{1}{c}{} & \multicolumn{6}{c}{emerging markets}\\
									\hline
									\multicolumn{1}{c}{} & EGX20    & BOVESPA    & JSE    & SSE  & CROBEX & XU100\\
									
									$\quad\Pi_{e_i}(s_i)$ & -0.68   & -0.59   & -0.56   & 0.29     & 0.56   &  0.63 \\      \end{tabular}
								
								\begin{tabular}{c c c c c c c c}
									\hline
									\multicolumn{1}{c}{} & \multicolumn{7}{c}{developed markets}\\
									\hline
									\multicolumn{1}{c}{} & CAC40    & FTSE100    &  NIKKEI    & NYSE  & BUX & SP500 & DAX \\                                       
									$\quad\Pi_{e_i}(s_i)$ & 1.09   & 1.18   & 1.22   & 1.22     & 1.24   &  1.34  &  1.36\\
									\hline                                 
								\end{tabular}
							\end{minipage}}
							\label{Tab_5}
						\end{table}

						%\clearpage
						
						%\begin{landscape}
						%\centering

						%\end{landscape}
						
						%\clearpage

						With this procedure we can examine the stock market time series in groups or individually, for any given SMI time series.
						
						\section{Conclusions}
						\label{end}
						
						In this paper we have analyzed spectral properties of time series of stock market indexes (SMIs) of developed, emerging, and underdeveloped (or transitional) market economies, in order to examine differences and similarities in their cyclical behavior, and to try to re-classify markets in our dataset according to the character of that behavior. We have used two different well established techniques of data analysis to obtain and verify our findings: the wavelet transformation (WT) spectral analysis and the time-dependent detrended moving average analysis (tdDMA). The combined use of these measures allowed us to identify a range of cycles universal to the SMI behavior across our dataset and to use the cyclic behavior to differentiate between levels of development of underlying SMI economies. This is the first study (to our knowledge) that has shown that cyclic behavior of SMI time series can be objectively differentiated for different SMI groups.
						
						We have found multiple peaks in wavelet spectra of all our SMI time series. Moreover, we have found all the peaks positioned at roughly the same times (or time intervals) in all our data, a finding that points to the similarity in seasonal behavior across different market economies in our dataset. We have identified what can be termed a working-week cycle  (or a 5-day peak), a one-week cycle (or a 7-day peak), a two-week cycle (or a 14-day peak), a monthly cycle (or a 30-day peak), a quarterly cycle (or a 90-day peak), a 4- to 5-month cycle (or a 150-day peak), a semi-annual cycle (or a 6- to 7-month peak), an annual cycle (or a 360-day peak), and a bi-annual (or a 600-days) multi-year cycle in our dataset. The dissimilarities between SMI records from the different economies that we have observed occur only in the lack of a spectral peak in some of the analyzed markets, or a slight lack of synchronization at a particular peak interval (peaks are not positioned at exactly the same time instances in all the SMI series analyzed). This prompted us to conclude that the seasonal behavior in different markets is probably a reflection of universality in market behavior, rather than a local characteristic of a particular economy. Given that financial markets are human-made complex systems, it is plausible to believe that our findings can be explained by the fact that business cycles are a reflection of common human working habits and behavior. Some authors find this commonality even desirable for the optimal functioning of a stock market, as was, for example, shown for the Euro monetary area \cite{ref13}. Some researchers, on the contrary, claim that these effects are not significant for the effectiveness of a stock market \cite{ref31}.
						
						In order to examine whether the observed seasonal adjustments in the behavior of stock markets could be used as indicators of the level of development or strength of the economy that underlies the specific market, we have performed a statistical analysis of the properties of wavelet spectra that characterize particular peak behaviors. We have statistically compared the relative energy content and the relative amplitude of each peak between the three groups of SMI series that we have analyzed - those belonging to developed economies, emerging economies and economically underdeveloped (or transitional) economies. We have found that the underdeveloped markets do not follow the same behavioral pattern as emerging or developed economies at the short time scales of days, weeks, and several months. Namely, their WT spectra show, in a statistically significant manner, less pronounced effects of fast (small time scale) cycles on the overall spectral behavior. In contrast, developed economies appear to even out all the cyclical (peak) effects in their WT spectra, or even to show a larger influence of the fast (small time scale) peak regions on their overall spectral behavior, while the emerging markets' spectra behave somewhere in the middle of these two cases. These observed differences could contribute to the variations in scaling behavior of markets, which has been reported previously \cite{ref18, ref25',ref33,ref34}. Namely, it has been shown that the economies of underdeveloped countries have WT spectra that show highly correlated long-range behavior, with the exponent $\beta > 0$ ($H > 0.5$), opposite to emerging and developed economies, which show uncorrelated or even slightly anti-correlated spectral behavior, with $\beta \leq 0$ ($H \leq 0.5$). The observed sensitivity of scaling exponents to the level of development of economies could be related to the findings we present here - to the relative influence of the small scale spectral peaks on the overall SMI spectral behavior.
						
						Finally, in this paper we propose a way to quantify the level of development of a stock market, based on the relative influence (or, in some cases, existence) of WT spectral peak intervals on the overall scaling behavior of SMI time series. In order to do that we have used the time-dependent Hurst exponent approach in a form of the tdDMA analysis, to calculate what we named the Development Index, which proved, at least in the case of our dataset, to be suitable to rank the SMI series in three distinct development groups. Further verification of this method remains open for future studies by us, or by other groups.

						% If in two-column mode, this environment will change to single-column format so that long equations can be displayed.
						%\begin{widetext}
						%$$\mbox{put long equation here}$$
						%\end{widetext}
						
						% Figures should be put into the text as floats.
						% Use the graphics or graphicx packages (distributed with LaTeX2e).
						% Put the label that you will use with \ref{} command in the braces of the \label{} command.
						%
						% \begin{figure}
						% \end{figure}
						
						% \begin{table}
						% \end{tabular}
						% \end{table}
						
						\section*{Acknowledgement}
							Acknowledgments: This work was supported by Serbian Ministry of Education, Science and Technological Development Research Grants No. 171015 and No. 174014. The work of Suzana Blesić has received funding from the European Union’s Horizon 2020 Research and Innovation Programme under the Marie Sk\l{}odowska-Curie Grant Agreement No. 701785.
					
						%\bibliography{your-bib-file}
						{}

					\end{document}